\documentclass[cameraready]{Interspeech}

\usepackage{amsmath,graphicx,hyperref}

\usepackage{cite}
\usepackage{newtxmath}
\usepackage{booktabs,multirow}

\usepackage[super]{nth}

\title{Voice Timbre Attribute Detection with Compact and Interpretable Training-Free Acoustic Parameters}

\author[affiliation={\ddagger}, orcid=0009-0005-9672-8147]{Aemon Yat Fei}{Chiu}
\author[affiliation={\ddagger}, orcid=0009-0009-5624-1704]{Yujia}{Xiao}
\author[affiliation={\ddagger}, orcid=0000-0003-2864-0475]{Qiuqiang}{Kong}
\author[affiliation={\ddagger,\diamond}, orcid=0000-0002-7089-3436]{Tan}{Lee}

\address{
    $^{\ddagger}$ Department of Electronic Engineering, The Chinese University of Hong Kong, Hong Kong \\
    $^{\diamond}$ School of Data Science, The Chinese University of Hong Kong, Shenzhen, China
}

\email{\{aemon.yf.chiu,yujiaxiao\}@link.cuhk.edu.hk, \{qqkong,tanlee\}@ee.cuhk.edu.hk}


\usepackage{comment}

\begin{document}

\maketitle

\begin{abstract}

    Voice timbre attribute detection (vTAD) is the task of determining the relative intensity of timbre attributes between speech utterances. Voice timbre is a crucial yet inherently complex component of speech perception. While deep neural network (DNN) embeddings perform well in speaker modelling, they often act as black-box representations with limited physical interpretability and high computational cost. In this work, a compact acoustic parameter set is investigated for vTAD. The set captures important acoustic measures and their temporal dynamics which are found to be crucial in the task. Despite its simplicity, the acoustic parameter set is competitive, outperforming conventional cepstral features and supervised DNN embeddings, and approaching state-of-the-art self-supervised models. Importantly, the studied set require no trainable parameters, incur negligible computation, and offer explicit interpretability for analysing physical traits behind human timbre perception.
    
\end{abstract}

\section{Introduction}


Voice timbre is an important component yet incomprehensible dimension of human speech for identifying and distinguishing speaking persons \cite{terasawa05_interspeech, kreiman2011foundations}. It is like the ``auditory face'' of a speaker \cite{Belin2004-sx}. Voice timbre conveys stable personal traits related to gender, age and physiological characteristics, and dynamic states like emotion and health \cite{laver1980phoetic}. According to \cite{asa_timbre}, timbre is defined as an attribute that ``\textit{enables a listener to judge that two \textbf{non-identical} sounds, similarly presented and having the same loudness, pitch, spatial location, and duration, are \textbf{dissimilar}}'', and is ``\textit{often specified by \textbf{qualitative adjectives} (e.g., bright or dull)}''. This highlights the complexity of timbre and the fact that voice timbre analysis relies on subjective natural language descriptions \cite{kreiman2011foundations, Kreiman2024-op}.

In \cite{he2025introducingvoicetimbreattribute}, a framework of voice timbre attribute detection (vTAD) was proposed for investigating and explaining the relationship between voice timbre and speech acoustics. In this framework, timbre attributes are defined by a group of verbal descriptors of voice timbre as listed in Table~\ref{tab:descriptor}. A vTAD system uses a Diff-Net to measure the comparative intensity of a timbre attribute between two speech utterances from different speakers. The ground-truth label is obtained from human perception \cite{vctk-rva}. The task of vTAD leverages a large-scale speech dataset which provides a direct mapping between speech utterances and subjective descriptions of voice timbre.  It aims to assess the performance of the latest speaker embedding models against human experts in perceiving voice timbre attributes.

\begin{figure}[b!]
  \centering
  \includegraphics[width=\linewidth]{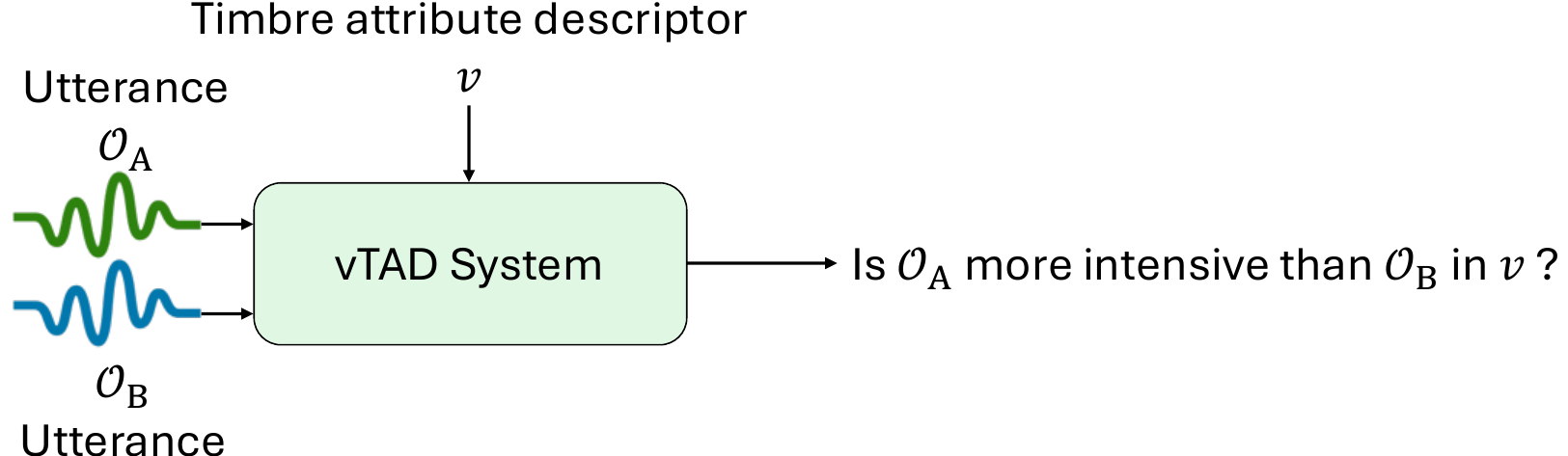}
  \caption{The definition of vTAD \cite{he2025introducingvoicetimbreattribute}.}
  
  \label{fig:vtad}

\end{figure}

The task of vTAD is similar to speaker verification (SV) as illustrated in Figure~\ref{fig:vtad}. General SV systems use speaker embeddings that entangle multiple factors of speech variation, e.g., content, timbre, prosody \cite{SpeechTripleNet,chiu2025largescaleprobinganalysisspeakerspecific}. vTAD focuses on evaluating how latent speaker embeddings specifically encode voice timbre information in speech. Despite that speaker embedding systems trained with large amount of data show high performance in vTAD \cite{chen2025voicetimbreattributedetection,ecapa-tdnn,naturalspeech3,cuhk-ee,song_vtad,deng_vtad,wu2025vtad}, these systems require excessive computational power for extracting high-dimensional embeddings and lack interpretability for meaningful voice analysis. The interpretability is vital for building reliable speech AI systems. It plays a significant role in real-world scenarios like forensics and legal settings.

\begin{table}[t!]
	\centering
		\caption{The descriptor set utilised in vTAD\cite{he2025introducingvoicetimbreattribute,vctk-rva}. \textbf{Proportion} indicates the percentage (\%) of each descriptor within the dataset. }
        \resizebox{\columnwidth}{!}{
	\begin{tabular}{lc lc}
		\toprule[1pt]
		
		\textbf{Descriptor} & \textbf{Proportion} & \textbf{Descriptor} & \textbf{Proportion} \\ 
		\cmidrule(lr){1-2} \cmidrule(lr){3-4}
		Bright            & 17.10      & Thin       & 13.03      \\
		Coarse       & 11.62      & Slim        & 11.31      \\
		Low         & 7.43       & Pure        & 5.48       \\
		Rich        & 4.71       & Magnetic    & 3.64       \\
		Muddy       & 3.59       & Hoarse      & 3.32       \\
		Round       & 2.48       & Flat        & 2.15       \\
		Shrill  (female-specific)      & 2.08       & Shrivelled   & 1.74       \\
		Muffled     & 1.44       & Soft        & 0.82       \\
		Transparent  & 0.66       & Husky (male-specific)      & 0.59       \\ 
		\bottomrule[1pt]
	\end{tabular}
}
	\label{tab:descriptor}
\end{table}

In the present study, a set of acoustic parameters is revisited and investigated for vTAD. The set contains 13 speech production related features and their temporal dynamics, which form a 26-dimensional vector. Compared with DNN learnt speaker embeddings, which are of hundreds or even thousands of dimensions and trained on large-scale speech data, the 26-dimension acoustic parameter set does not require GPU acceleration for feature extraction, yet achieves comparable performance on the task and offers explicit interpretability. Analysis of experimental results reveals that the temporal dynamics of speech play an important role in distinguishing timbre attributes. The contributions of individual acoustic features are analysed to provide insights into interpretable voice timbre analysis and speaker embedding design.

\section{Related Work}

The research on timbre initially focused on its physical correlates \cite{1875,1926}. Later works shifted attention to the perceptual aspects of timbre \cite{mcadams1993, HANDEL1995425}. A clear and comprehensive definition of timbre remains elusive \cite{wei2022}, despite attempts to construct timbre spaces for speech \cite{terasawa05_interspeech} and to develop acoustic parameter sets for timbre categorisation \cite{vankova14_speechprosody,acousticvoicevariation,kreiman2021,Kreiman2024-op}.

Current approaches to vTAD are dominated by data-driven, DNN-based speaker embedding models and classifiers \cite{chen2025voicetimbreattributedetection}. Chiu et al. \cite{cuhk-ee} applied WavLM-Large \cite{wavlm} as the speaker encoder. Song et al. \cite{song_vtad} investigated an improved method of speaker embedding fusion. Other studies \cite{deng_vtad,wu2025vtad} focused on improving the downstream Diff-Net for better classification performance. High-dimension embeddings learnt from large amount of speech data lack interpretability. They perform well in differentiating voice timbres using black-box model analysis and  do not give any meaningful insight on why the timbres are similar or distinct.


With the rapid development of AI and DNN in recent decades, various speaker embeddings and representations have been proposed to represent unique speakers in the latent space \cite{snyder17_interspeech,x-vector,ecapa-tdnn,r-vector,cam++,wavlm}. The learnt embeddings typically entangle multiple speaker-related components including timbre, prosody, and accent \cite{luu22_interspeech,chiu2025largescaleprobinganalysisspeakerspecific}. There were attempts to explicitly disentangle these components and explicitly model the timbre of human speech \cite{SpeechTripleNet,naturalspeech3}.


Liu et al. \cite{liu24q_interspeech} investigated the consistency among human perception, DNN based SV systems, and an acoustic parameter set for voice similarity measurement. It was found that the acoustic parameters moderately align with human perception making binary same/different judgements, but demonstrate limited alignment with human judgement in rating speaker similarity. SV systems align with human perception only at a broad and gross-phonetic level. Both automatic systems fail to fully capture the fine phonetic details of human auditory judgments, even though the two groups of features show high internal correlation \cite{liu24q_interspeech}.


The present work shifts the focus from general SV to the more focused task of vTAD. A large dataset with fine-grained expert annotations and controlled natural language framework for timbre attribute description are employed. In this study, expert annotations are treated as the ground truth. The computed features are evaluated by comparing their outputs against expert labels.

\begin{figure}[b!]
  \centering
  \includegraphics[width=\linewidth]{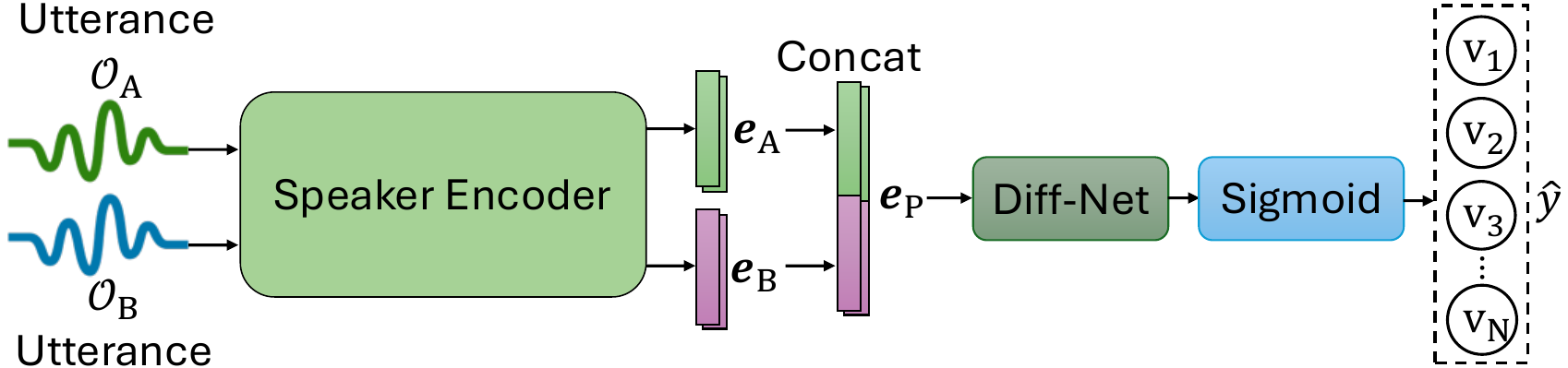}
  \caption{The overall workflow of vTAD \cite{he2025introducingvoicetimbreattribute}.}
  \label{fig:vtad_task}
\end{figure}

\section{Acoustic Parameters for vTAD}

\subsection{Task definition}

In the vTAD task, a pair of speech utterances ${\mathcal O}_{\rm A}$ and ${\mathcal O}_{\rm B}$ from speaker A and speaker B are fed into the detection system, together with a timbre descriptor $v$ $\in$ $V$. The system generates a prediction score in the range of $[0,1]$. A score close to $0$ indicates a high confidence that ${\mathcal O}_{\rm A}$ is more intense in the timbre attribute $v$, and vice versa \cite{he2025introducingvoicetimbreattribute,sheng2025voicetimbreattributedetection}. Figure~\ref{fig:vtad_task} illustrates the detailed workflow of a vTAD system. Speaker embeddings ${e}_{\rm A}$ and ${e}_{\rm B}$ are extracted and concatenated to form ${e}_{\rm P}$ for both ${\mathcal O}_{\rm A}$ and ${\mathcal O}_{\rm B}$ through a frozen speaker encoder, and passed on to the Diff-Net to obtain the predicted scores $\hat y$ for timbre attributes $v_1$ to $v_N \in V$.

\subsection{Acoustic parameters}


In the proposed system, the speaker encoder extracts 13 base acoustic features and their respective coefficients of variation (CoVs), adopted from \cite{acousticvoicevariation,kreiman2021,Kreiman2024-op}. These acoustic parameters are summarized in \cite{Kreiman2024-op,acousticvoicevariation}. The base parameters consist of fundamental frequency ($F_0$), the first four formant frequencies ($F_1$, $F_2$, $F_3$, and $F_4$), formant dispersion, four harmonic spectral shape measures ($H_1^*-H_2^*$, $H_2^*-H_4^*$, $H_4^*-H_\text{2kHz}^*$, and $H_\text{2kHz}^*-H_\text{5kHz}$\let\thefootnote\relax\footnotetext{* Correction for the influence of formants on harmonic amplitudes.}), and three inharmonic source metrics: cepstral peak prominence (CPP), root mean square (RMS) energy, and sub-harmonic-to-harmonic ratio (SHR). These 13 parameters and their CoVs form a 26-dimensional feature vector.

While the core acoustic definitions summarised in \cite{acousticvoicevariation} are retained, the extraction methodology in this work includes several modifications for robustness and dynamic alignment. In this study, the Praat-Parselmouth tool \cite{parselmouth} is used to extract the acoustic parameters. A 10 ms time step for raw acoustic measurements is implemented to capture the highly dynamic voice timbre. $F_0$ is extracted via autocorrelation with a pitch floor of 75 Hz. Formants and their bandwidths are estimated using Burg's algorithm, capped at a maximum of five formants up to 5.5 kHz. Formant dispersion is defined computationally as the frequency difference between $F_4$ and $F_1$ divided by three.



To calculate spectral and energy features, a static 40 ms analysis window is utilised for each voiced frame. Within these slices, RMS energy is calculated, and spectral tilt measures are extracted and subsequently corrected for the overlapping influence of formant frequencies and bandwidths. SHR is derived from the magnitude spectrum of the slice, and CPP is extracted via a robust parabolic power cepstrogram approach. Both SHR and CPP extractions utilise Praat-Parselmouth's \cite{parselmouth} built-in functions.

Following the extraction of these raw time-series metrics, the global mean and global CoV across all valid voiced frames are computed directly to construct the final compact 26-dimensional utterance-level representation per audio file.

\subsection{Downstream classifier}
\label{sec:3.2}
A simple Diff-Net, which comprises two fully-connected (FC) layers with a batch normalisation (BN) layer, a ReLU activation function, and a dropout layer in between, is implemented as the downstream classifier and trained to conduct vTAD.

\section{Experimental Settings}

\subsection{Dataset}

The VCTK-RVA dataset \cite{vctk-rva}, which originates from the VCTK dataset \cite{vctk} and is enriched with human annotation of timbre attribute intensity by speech experts, is utilised in this study. Within this dataset, 6,038 same-gender speaker pairs involving 101 speakers (40 male and 61 female) are annotated. Each speaker pair possesses intensity annotations for one to three timbre attributes \cite{he2025introducingvoicetimbreattribute,vctk-rva}.

A training set and a test set are established to train the Diff-Net and evaluate the vTAD system. Originally, the training set contained 136,320 utterances from 78 speakers (including 29 male speakers and 49 female speakers) \cite{he2025introducingvoicetimbreattribute}. In this work, utterance pairs containing the sample \texttt{p341/p341\_101}, which is a silent file, are removed from the training set, resulting in 136,316 valid pairs. The \textit{unseen} test set from \cite{he2025introducingvoicetimbreattribute} is adopted, ensuring no speaker overlap with the training set. The test set contains 91,600 utterance pairs from 23 speakers (11 male and 12 female).

For data pre-processing, the audio files are retrieved directly from \cite{vctk} utilising only the \texttt{\_mic1.flac} files. The data are converted from FLAC to WAV format and downsampled from 48 kHz to 16 kHz without further modification, aligning strictly with the settings in \cite{he2025introducingvoicetimbreattribute,vctk-rva}.

\subsection{Baseline systems}
Five speaker embeddings and cepstral features are evaluated as baseline speaker encoders in this study: ECAPA-TDNN \cite{ecapa-tdnn}, FA-Codec \cite{naturalspeech3}, Mel-Frequency Cepstral Coefficients (MFCCs), Linear Frequency Coefficients (LFC) \cite{terasawa05_interspeech}, and WavLM \cite{wavlm}.

\begin{itemize}
\item ECAPA-TDNN \cite{ecapa-tdnn} is a popular supervised speaker embedding model trained for SV. FA-Codec is the speaker encoder in NaturalSpeech 3 \cite{naturalspeech3}, a zero-shot text-to-speech synthesis system capable of replicating a speaker's voice with high naturalness and similarity. The 1024-Channel ECAPA-TDNN \cite{ecapa-tdnn} and the timbre branch of FA-Codec \cite{naturalspeech3} serve as the two primary baselines for the vTAD task \cite{he2025introducingvoicetimbreattribute}.

\item MFCCs are commonly used audio signal features widely adopted in various downstream tasks, including speech and speaker recognition. LFC is a ``strawman'' representation that replaces the Mel-frequency scale in MFCCs in \cite{terasawa05_interspeech}. Both MFCC and LFC are tested in an early investigation of latent features for speech timbre and are evaluated here using implementations that follow \cite{terasawa05_interspeech}.

\item WavLM \cite{wavlm} is a large speech self-supervised learning model comprising a stack of Transformers \cite{attention} trained on 960 to 94,000 hours of speech data. Chiu et al. \cite{cuhk-ee} utilise attentive statistic pooling (ASTP) \cite{astp} to aggregate the outputs of all Transformer layers from WavLM-Large \cite{wavlm} to obtain the final speaker embedding.
\end{itemize}


The Diff-Net for WavLM models contains four blocks comprising FC, BN, ReLU, and dropout layers before the final FC layer \cite{cuhk-ee}, while other systems strictly observe the setting outlined in Section~\ref{sec:3.2}.

\section{Results and Analysis}

\subsection{Performance analysis}

The performance of the acoustic parameter set against several baseline systems on the vTAD task, measured by Acc and EER, is shown in Table~\ref{tab:baseline_results}. WavLM-Large with ASTP-\textbf{L}, a multi-layer output aggregation method, achieves state-of-the-art (SOTA) performance. The 26-dimensional acoustic parameter set achieves an Acc of 82.87\% and an EER of 17.21\%. It outperforms the final representations of the WavLM models (WavLM-Base, WavLM-Base+, and WavLM-Large), cepstral features MFCCs and their variant LFC, and dedicated supervised speaker embedding models, i.e., ECAPA-TDNN and the timbre embedding of FA-Codec. Furthermore, its performance is very close to the SOTA.

The results for speaker embeddings are consistent with prior findings \cite{he2025introducingvoicetimbreattribute} where FA-Codec exhibits better generalisation ability on unseen data, while ECAPA-TDNN has better performance on seen test data where the speakers already appear in the training set. The comparison between two common audio features, MFCC and LFC, however, shows a different result from the prior study \cite{terasawa05_interspeech}, which argues that MFCC is a better feature for perceptual timbre analysis. Table~\ref{tab:baseline_results} shows that the LFC performs significantly better than MFCC in the vTAD task. Because the Mel scale heavily compresses high-frequency information, it may obscure the high-frequency inharmonic energy and spectral noise that are critical for perceiving specific timbre attributes \cite{acousticvoicevariation}. By preserving linear resolution across the spectrum, LFC captures these high-frequency details more effectively, aligning with findings that linear-frequency scales are more robust for speaker recognition tasks \cite{lfcc-mfcc}.


\begin{table}[t!]
	\centering
	\caption{Performance of various systems on vTAD measured by accuracy (Acc) and equal error rate (EER).}
    \scalebox{1}{
	\begin{tabular}{lll}
        \toprule
        \textbf{Model} & \textbf{Acc (\%)}↑ & \textbf{EER (\%)}↓ \\ 
		\midrule
		ECAPA-TDNN  & 70.37  & 28.67 \\
		FA-Codec & 79.32  & 20.60 \\
        \midrule
        MFCC & 68.72 & 31.15 \\
        LFC & 80.32 & 19.41 \\
        \midrule
        WavLM-Base & 75.76 & 23.89 \\
        WavLM-Base+ & 78.02 & 22.23 \\
        WavLM-Large & 80.08 & 19.21 \\
        \quad w/ ASTP-\textbf{\textit{L}} \cite{cuhk-ee}  & \textbf{83.13} & \textbf{16.87} \\
        \midrule
        Acoustic Parameters & 82.87 & 17.21 \\
		\bottomrule
	\end{tabular}}
	\label{tab:baseline_results}

\end{table}

\begin{figure*}[ht]
  \centering
  \includegraphics[width=\linewidth]{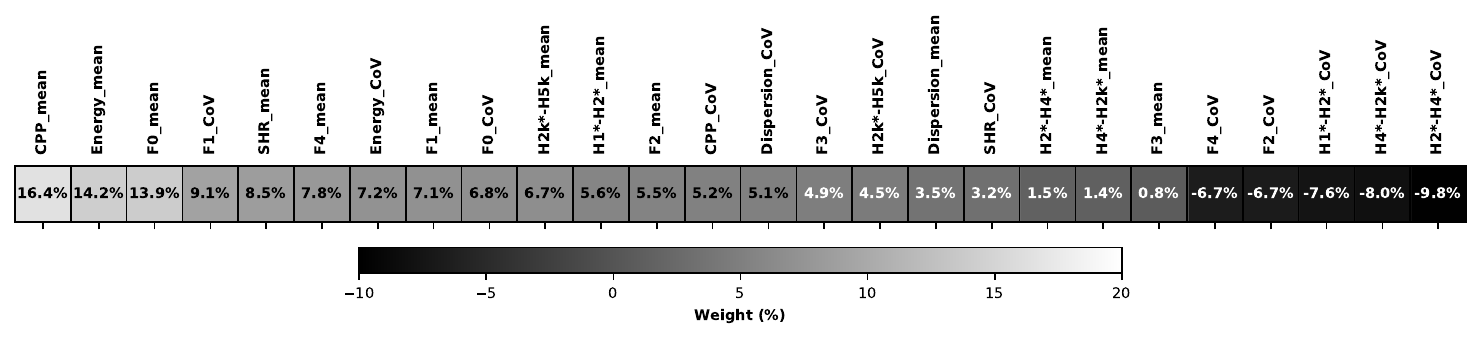}
  \caption{Feature weights in the Diff-Net using the acoustic parameter set.}
  \label{fig:feature_importance}
\end{figure*}

Table~\ref{tab:baseline_results} also reveals that trained models can benefit from larger training data sizes and model scales, as seen from the performance differences among WavLM-Base, WavLM-Base+, and WavLM-Large. WavLM-Base consists of 12 Transformer layers with 768 hidden states and is trained on 960 hours of speech data. WavLM-Base+ shares the same model architecture with WavLM-Base but is trained on 94,000 hours of speech data, similar to WavLM-Large that consists of 24 Transformer layers with 1,024 hidden states. It is also shown that utilising the intermediate representations of WavLM-Large can further significantly improve performance.

\subsection{Interpretability}

The fundamental advantage of the acoustic parameter set lies in its physical interpretability. Figure~\ref{fig:feature_importance} illustrates the feature importance weights derived directly from the trained Diff-Net with an added simple learnable weight layer. The mean values of CPP, energy, $F_0$, SHR, and the CoV of $F_1$ serve as the most significant positive indicators for distinguishing timbre attributes. $F_0$ represents the actual vibratory rate of the vocal folds, while CPP serves as a direct metric for the degree of periodicity and harmonic richness, both of which are foundational to perceived voice timbre. The large weight of SHR mean demonstrates that the presence of period doubling can also be a significant indicator for vTAD. Formant frequency $F_1$, which is generally more related to vowel identity, is also shown to play an important role in timbre detection. Dynamic variations in the spectral slopes (e.g., the CoVs of $H_2^*-H_4^*$, $H_4^*-H_\text{2kHz}^*$, and $H_1^*-H_2^*$), which are related to breathiness or auditory brightness, act as the most important negative weights in the network. This indicates the discriminative ability of the temporal variability of high-frequency inharmonic energy in separating the perceptual profiles of different speakers. The results are consistent with \cite{acousticvoicevariation} that the variability in high-frequency spectral shape and spectral noise are significant in distinguishing speaker timbre. However, \cite{acousticvoicevariation} claims that $F_0$ has little importance in the voice space, whereas the present work finds that $F_0$ is a primary distinguishing feature, aligning with recent speaker modelling studies that generally treat $F_0$ as a fundamental speaker-specific attribute \cite{Kreiman2024-op,chiu2025largescaleprobinganalysisspeakerspecific,textrolspeech,speechcraft}.

Traditional baseline cepstral features like MFCC and LFC coefficients are derived via a discrete cosine transform applied to filterbanks. This decorrelates the spectral envelope into mathematical basis functions that cannot be mapped back to interpretable physical traits. High-dimensional DNN-based models, including ECAPA-TDNN, FA-Codec, and WavLM, operate as black boxes and entangle various speech components into an abstract latent space. Consequently, their output embeddings generally lack interpretability. Crucially, these speaker embedding models generally perform frame-averaging, which eliminates the temporal dynamics that are shown to be important in distinguishing timbre attributes. Even when employing widely used techniques like ASTP that do capture variance, this dynamic information is projected into an abstract latent space where it becomes heavily entangled with the mean

\subsection{Efficiency}

Beyond offering interpretability, the acoustic parameter set is also highly computationally efficient and requires no GPUs. As shown in Table~\ref{tab:extract_flops_params}, the acoustic parameter set extraction process requires zero trainable parameters and operates at a highly efficient 17.85 M floating-point operations per second (FLOPs) for one second of speech to output a 26-dimensional vector per utterance. In contrast, trained DNN models require 10 to 300 M trainable parameters and 80 M to 25 G FLOPs per second of speech, outputting embeddings ranging from 192 to frame-by-1024 dimensional (which are further expanded by aggregating representations across all 25 layers of the WavLM-Large to achieve optimal performance). The difference in embedding size indicates that training the downstream Diff-Net for the acoustic parameter set is also more efficient than DNN-based methods, as demonstrated in Table~\ref{tab:train_flops_params}. While MFCC and LFC extraction and downstream training are marginally cheaper, they suffer from lower performance and relative lack of interpretability.

\begin{table}[t!]
	\centering
	\caption{Computational costs for speaker embedding extraction process of each system measured by model parameters (params), FLOPs per second (sec), and dimension (dim) per utterance (uttr).}
\resizebox{\columnwidth}{!}{
    \begin{tabular}{llll}
        \toprule
        \textbf{Model} & \textbf{Params} & \textbf{FLOPs per sec} & \textbf{Dim per uttr} \\ 
		\midrule
		ECAPA-TDNN & 17.24 M & 84.33 M  & 192 \\
		FA-Codec & 103.61 M & 3.57G & 256 \\
        \midrule
        MFCC & 0 & 5.41 M & 13 \\
        LFC & 0 & 4.46 M & 13 \\
        \midrule
        WavLM-Base & 94.70 M & 7.25 G & 13 $\times$ [\textbf{\textit{F}}, 768] \\
        WavLM-Base+ & 94.70 M & 7.25 G & 13 $\times$ [\textbf{\textit{F}}, 768] \\
        WavLM-Large & 316.62 M & 25.88 G & 25 $\times$ [\textbf{\textit{F}}, 1,024] \\
        \midrule
        Acoustic Parameters & 0 & 17.85 M & 26 \\
		\bottomrule
	\end{tabular}}
	\label{tab:extract_flops_params}
\end{table}

\begin{table}[t!]
	\centering
	\caption{Computational costs for the Diff-Net training process of each system measured by model parameters (params) and FLOPs per utterance (uttr) pair.}
    \scalebox{1}{
    \begin{tabular}{lll}
        \toprule
        \textbf{Model} & \textbf{Params} & \textbf{FLOPs per uttr pair} \\ 
		\midrule
		ECAPA-TDNN & 70.31 k & 140.80 k \\
		FA-Codec & 53.92 k & 108.03 k \\
        \midrule
        MFCC & 7.84 k & 15.36 k \\
        LFC & 7.84 k & 15.36 k \\
        \midrule
        WavLM-Base & 1.03 M & 2.06 M\\
        WavLM-Base+ & 1.03 M & 2.06 M \\
        WavLM-Large & 1.29 M & 2.59 M \\
        \quad w/ ASTP-\textbf{\textit{L}}  \cite{cuhk-ee} & 10.74 M & 844.36 M \\
        \midrule
        Acoustic Parameters & 11.17 k & 22.02 k \\
		\bottomrule
	\end{tabular}
    }
	\label{tab:train_flops_params}
\end{table}

\section{Conclusions}

This study demonstrates that a compact and interpretable 26-dimensional set of acoustic parameters can effectively capture the elusive nature of voice timbre, acting as a powerful candidate against complex high-dimensional DNN-based embeddings for the vTAD task. The physically grounded set achieves 82.87\% accuracy on the voice timbre attribute detection task, outperforming widely used supervised speaker embeddings and traditional cepstral features. Furthermore, it achieves performance comparable to the state-of-the-art WavLM model while offering improved interpretability and requiring significantly lower computational resources. Temporal variations in speech are found to be critical for timbre perception but are not explicitly modelled in DNN speaker embeddings. The findings suggest that integrating interpretable acoustic knowledge into modern artificial intelligence systems represents a promising direction for explainable and efficient speaker trait analysis.

\section{Generative AI Use Disclosure}

The authors disclose that generative AI tools were utilised solely for the purpose of editing and polishing this manuscript, including correcting grammatical errors, refining sentence structures, and improving the overall clarity of the English text. All (co-)authors take full responsibility and accountability for the original research, data analysis, and technical content presented in the manuscript. No generative AI tool was used to produce any significant part of the manuscript.

\bibliographystyle{IEEEtran}
\bibliography{main}

\end{document}